# Robust, directed assembly of fluorescent nanodiamonds.

Mehran Kianinia[1], Andreas W. Schell[2], Olga Shimoni[1], Steven J. Randolph[3], Milos Toth[1], Igor Aharonovich[1*] and Charlene J. Lobo[1*]

1. School of Mathematical and Physical Sciences, University of Technology, Sydney, P.O. Box 123, Broadway, New South Wales 2007, Australia.

2. Department of Electronic Science and Engineering, Kyoto University, Kyoto daigaku-katsura, Nishikyo-ku, Kyoto, Japan

3. FEI Company, 5350 NE Dawson Creek Dr., Hillsboro, OR 97124, USA

**Abstract**

*Arrays of fluorescent nanoparticles are highly sought after for applications in sensing and nanophotonics. Here we present a simple and robust method of assembling fluorescent nanodiamonds into macroscopic arrays. Remarkably, the yield of this directed assembly process is greater than 90% and the assembled patterns withstand ultra-sonication for more than three hours. The assembly process is based on covalent bonding of carboxyl to amine functional carbon seeds and is applicable to any material, and to non-planar surfaces. Our results pave the way to directed assembly of sensing and nanophotonics devices.*

Assembling fluorescent nanoparticles into macroscopic arrays is required for many applications spanning sensing[1], photonics, plasmonics and quantum information processing[2-4]. To achieve this goal several top down techniques, including lithography[5,6] or dip-pen techniques[7,8], as well as bottom up methods using patterned self-assembled monolayers[9] or electrostatic self-assembly[10,11] have been developed. While these methods are capable of high resolution patterning of nanoparticle arrays, the assembled components are only weakly bonded to the substrate and cannot undergo further wet chemistry processing steps (eg sonication) or subsequent lithography. Such processing is often required for device applications where the fluorescent nanoparticles act as active components in microfluidic devices[12,13], as sensing probes[14] or photon sources in which they are coupled to plasmonic structures or other optical elements[15-17].

In particular, there is a great interest in controlling and positioning fluorescent nanodiamonds that host nitrogen vacancy (NV⁻) defects, which can then be employed as nanoscale sensors for detection and imaging of weak magnetic fields[1, 18, 19], thermal imaging or thermometry[20-22] and quantum measurements[23-25]. Moreover, there is a great interest in assembling arrays of nanodiamonds[26, 27] that can subsequently be used to couple to plasmonic waveguides to realize quantum plasmonics circuitry[17, 28, 29]. However, to date there is no robust method of accurately positioning nanodiamonds in arrays that can be subjected to further processing steps that are needed for device fabrication.

Here we realize a facile, robust method for high resolution self-assembly of nanodiamonds which enables their use in sensing, photonic and quantum devices. We employ nanoscale seeds that are fabricated in a single step by a mask-free electron beam induced deposition (EBID) technique[30], terminate the seeds with amine groups, and self-assemble nanodiamonds into arrays defined by the seed positions. The technique is not limited to any specific substrate and can be used to position nanodiamonds on arbitrary materials and non-planar surfaces. Finally, the technique offers high stability, which we demonstrate by subjecting the fabricated nanodiamond arrays to multiple sonication steps of up to 12 hours total duration.

The nanodiamond patterning process is illustrated in Fig. 1a and described in detail in Supplementary section S1. In step 1, electron beam induced deposition (EBID) in a variable pressure SEM was used to fabricate nanoscale carbonaceous seeds using the organic precursor naphthalene ($C_{10}H_8$). Carbon seeds were deposited in arrays using a stationary defocused electron beam (15 keV, 300 pA, 30s), resulting in disks of approximately 90 nm diameter and 20 nm height, as seen in Fig 1b, c. In step 2, the EBID seeds were amine-functionalized by 45s exposure to an ammonia plasma generated in a Reactive Ion Etching (RIE) system operating at 100W and 6 Pa $NH_3$. These conditions have been reported to produce the highest concentration of $NH_2$ groups in the plasma[31]. The extent and nature of amine groups created in the surface carbon was assessed by X-ray photoelectron spectroscopy (Supplementary section S2). The final step involves covalent attachment of 35 nm nanodiamonds[32] to the EBID seeds using 1-Ethyl-3-(3-dimethylaminopropyl)carbodiimide (EDC). The presence of dangling bonds at the nanodiamond surface allows them to be functionalized with a variety of ligands[33, 34]. The surfaces of oxidized nanodiamonds are terminated with carboxylic acid (-COOH) groups, enabling their attachment to amine-terminated surfaces through carbodiimide coupling chemistry. Conjugation was achieved by immersing the substrate that contained the EBID seeds in an aqueous solution of EDC and varying concentrations of nanodiamonds for 6 hours. Samples were then washed with DI water and dried in $N_2$. Figure 1d shows the resulting patterned array of nanodiamonds (the inset shows a high resolution image of a single EBID seed with several nanodiamond crystals attached to it).

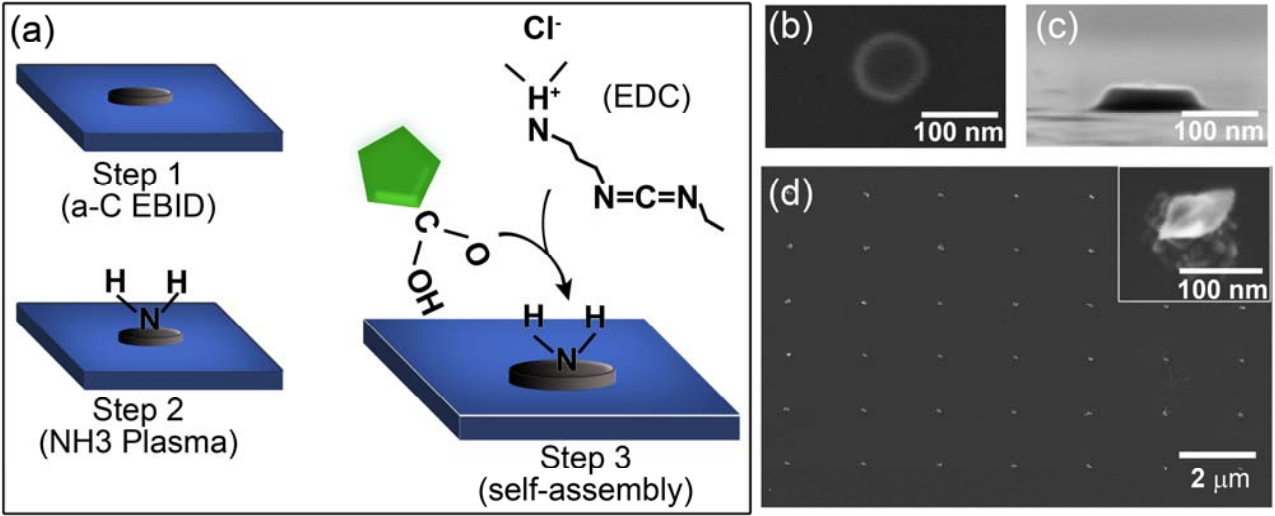

**Figure 1. Self-assembly of nanodiamonds on electron beam deposited carbon seeds. (a) Schematic of the process. (b, c) Plan and side view SEM images of amorphous carbon disks deposited by EBID. (d) SEM image of patterned area after nanodiamond attachment. The inset in (d) is a high magnification SEM image of an EBID seed with attached nanodiamonds.**

To ascertain the nanodiamond attachment yield and selectivity of the technique, we employ confocal microscopy to obtain photoluminescence maps and spectra of the fabricated arrays. For the optical measurements, we employed a home built confocal microscope with a high numerical aperture objective (100 x, 0.9 N.A), used for both excitation and collection of the emitted light. A 532 nm continuous wave laser was used for excitation, and all measurements were done at room temperature under ambient conditions. Figure 2a shows a SEM image of a nanodiamond array and Figure 2b shows a confocal map of the same array. The bright fluorescent spots correspond to the emission from nitrogen vacancy (NV$^-$) defects in the nanodiamonds. Figure 2c shows the spectrum recorded from each spot, demonstrating that 32 out of 35 locations have the characteristic emission from the NV$^-$ centres, equating to a 92% yield for the attachment process. Note that no nanodiamonds were attached in between the EBID seeds, giving the technique 100% selectivity.

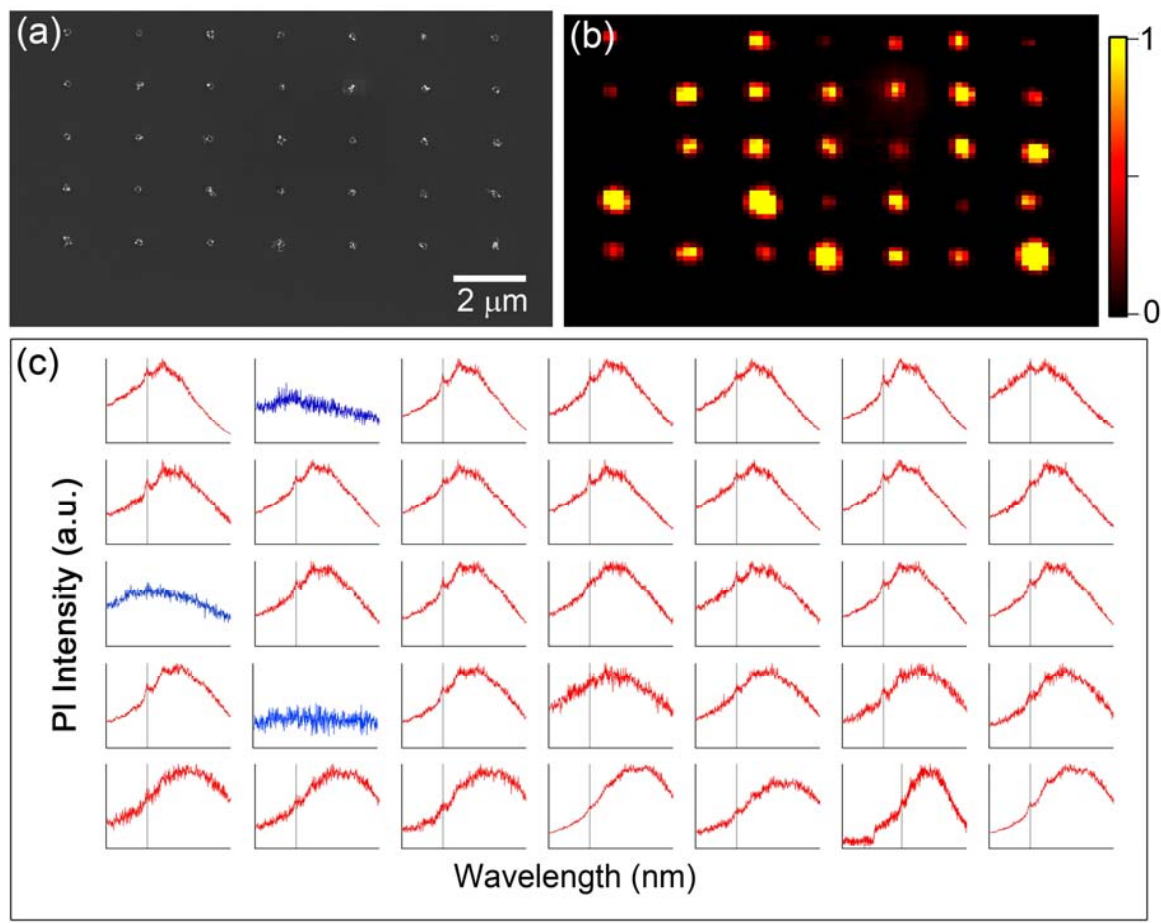

**Figure 2. Characterization of attachment yield and selectivity. (a) SEM image of a nanodiamond pattern. (b) Colour-coded confocal photoluminescence map of the same nanodiamond pattern. (c) Corresponding PL spectra from each individual spot showing the NV$^-$ emission. The NV$^-$ zero phonon line is marked with a vertical gray line for clarity. Only three locations in the patterned array do not show the NV$^-$ ZPL (blue curves).**

We now study the effect of the nanodiamond concentration on the attachment yield. Figure 3a shows a clear dependence of the yield on the initial concentration of fluorescent nanodiamonds. Increasing the nanodiamond concentration results in improved attachment yield. At a nanodiamond concentration of 2.5 μg/ml the attachment yield was smaller than 5%, increasing to 50 % at a concentration of 12.5 μg/ml. The optimum concentration was found to be 25 μg/ml, resulting in greater than 92% attachment, with only a few spots having no nanodiamonds. A higher concentration of nanodiamonds resulted in agglomeration with no increase in yield. The probabilities were deduced by analysis of confocal maps recorded under the same conditions as in Fig. 2 (and shown for each data point in Fig. 3). For all experiments, the ratio of EDC:nanodiamonds in the solution was fixed at 10:1.

The effectiveness of our method relies on covalent bonding between the amine and carboxyl functional groups on amorphous carbon and nanodiamond surfaces in the presence of EDC. We therefore expected the attached nanodiamonds to withstand further processing and treatment, as is required for many device applications. To examine the robustness of the technique, the assembled nanodiamonds were sonicated in a powerful ultrasonic bath (Bransonic 185 Watt Ultrasonic cleaner 221). Figure 3b shows the remaining nanodiamonds after multi-step ultra-sonication for up to 12 hours. After three hours of sonication, all the nanodiamonds that were initially assembled were still attached to the substrate. Even after 12 hours, more than 90% of the self-assembled nanodiamonds remained on the surface, proving the unprecedented robustness of the assembly technique.

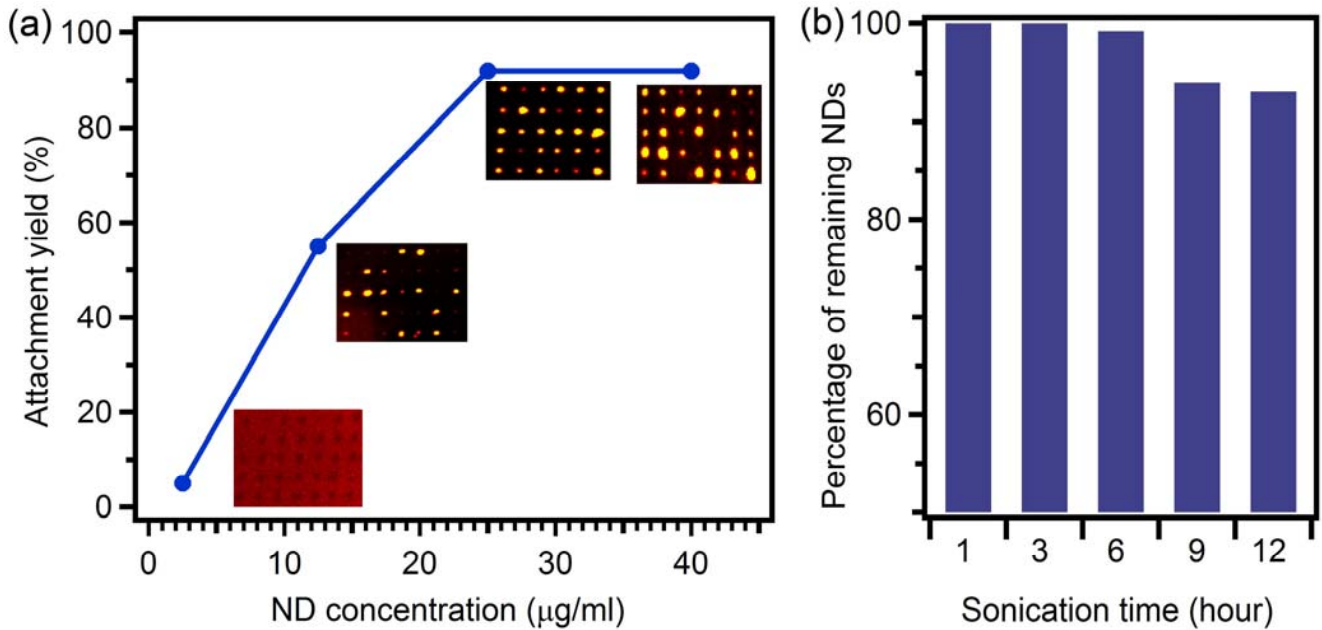

**Figure 3. (a) Effect of nanodiamond concentration on the attachment yield. The yields were calculated from confocal maps shown next to each data point. (b) Robustness of nanodiamond array under sonication.**

Finally, we demonstrate that the formed array can be used as a high-resolution magnetic field sensor, where each element in the array can serve as an individual pixel. Here, a microwave is guided through a 30 µm wire and the NV⁻ spin states are read out optically (so-called optically detected magnetic resonance, ODMR, described in Supplementary section S3) (Fig 4a)[35]. Such sensing of magnetic fields in ambient environments is one of the most prominent applications of the NV⁻centre. Figure 4b shows a confocal map of the array of NV⁻ centers. Figure 4 (c-e) shows three examples of optically detected magnetic resonance (ODMR) from randomly selected pixels in the array (marked with green circles). The red curves are nearly identical for all pixels and correspond to zero magnetic field. The green and the blue curves show the ODMR under 1 mT and 3 mT magnetic field. Each pixel shows a distinct ODMR signal that can then be used to deduce the local magnetic field in the proximity of the pixel. Note that since the measurement is done with an ensemble of nanodiamonds, the signal is broadened, with each dip comprising several lines from different NV⁻ centers at each spot in the array. This technique is ideal to test for local absorption of metal nanoparticles or the presence of foreign para- and ferromagnetic metals. In principle, our technique can also be applied to nanodiamonds with single emitters, which are advantageous for quantum photonic applications.

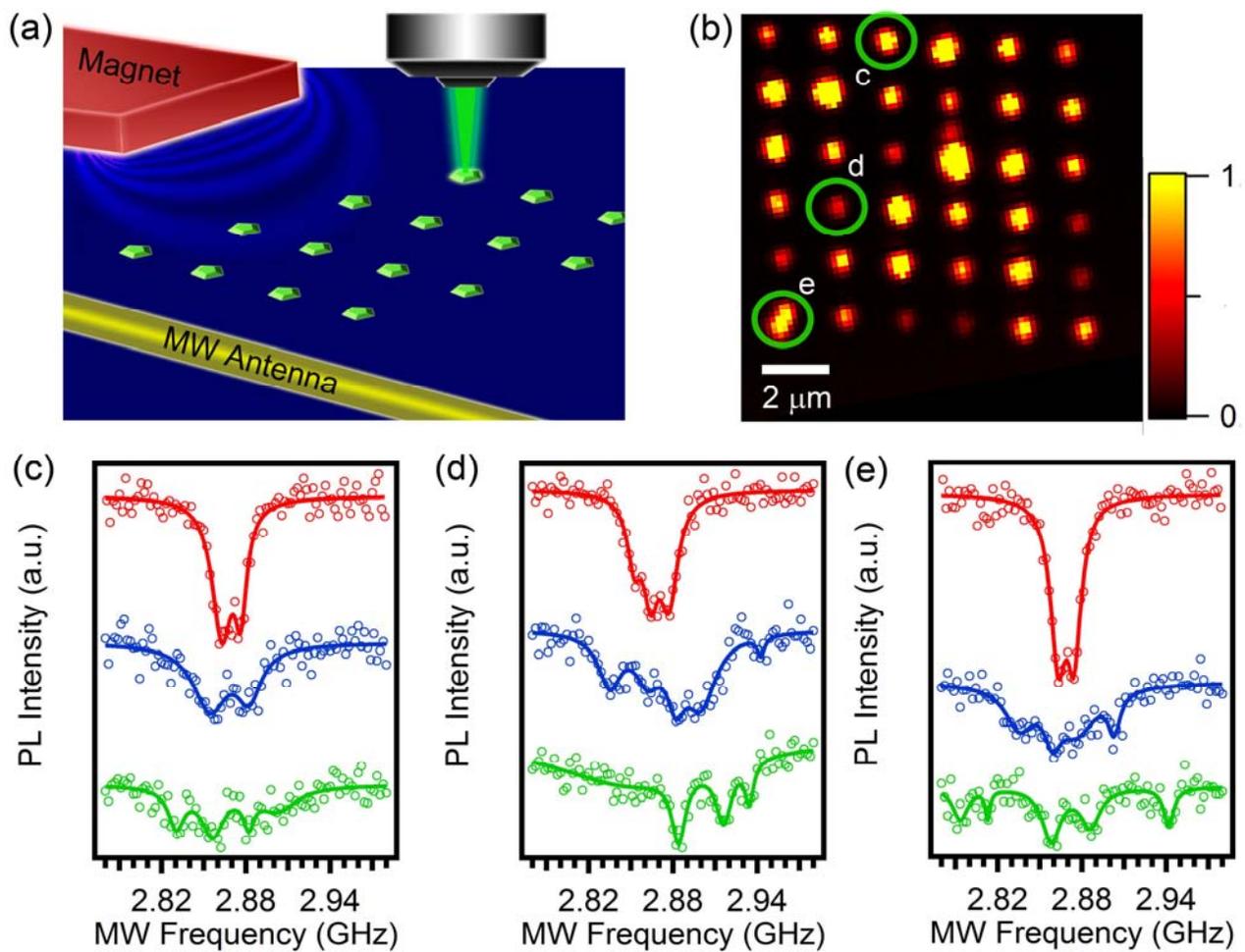

**Figure 4.** Demonstration of the nanodiamond array operating as a magnetic field sensor. (a) A microwave is guided through a 30 μm wire and electron spins are read out optically (b) confocal map of the nanodiamonds assembled into an array. (c-e) optically detected magnetic resonance (ODMR) signal recorded from different elements (pixels) in the array – as shown by green circles. Red curve is ODMR under zero magnetic field while blue and green curves correspond to 1 and 3 mT fields, respectively.

In conclusion we have developed a facile, generic technique for directed assembly of fluorescent nanodiamonds into robust arrays. The assembly technique has greater than 90 percent efficiency. Moreover, the nanodiamonds are covalently bonded and stay in their positions even after repeated ultrasonication treatments, making the technique very attractive for practical device applications. Finally, we have performed a proof of principle sensing measurement of various magnetic fields to show that each pixel in the array can be used as an independent magnetic field sensor. Our method paves the way to realization of scalable platforms for sensing or integrated quantum photonics, where there is a real need for large area assembly of fluorescent nanoparticles. It is important to note that while applied to nanodiamonds in this work, the technique is versatile and can be used to assemble other nanoparticles on arbitrary surfaces.

AUTHOR INFORMATION

***Corresponding Authors:** Charlene Lobo, charlene.lobo@uts.edu.au, Igor Aharonovich, igor.aharonovich@uts.edu.au.

**Author Contributions**

The work has been done by the contribution of all authors. All authors have discussed the results and given approval to the final version of the manuscript.

ACKNOWLEDGMENTS

This work was funded by FEI Company and the Australian Research Council. The authors thank Hugh Mackay for his contribution with EBID of carbon seeds and Ari Bendavid for performing XPS measurements.